\documentclass[11pt]{article}
\usepackage{amsmath, amsthm, amssymb}
\usepackage{graphicx}

\parskip \baselineskip
\newtheorem{proposition}{Proposition}[section]

\begin{document}

{\huge \bf{Marshall-Olkin Extended Zipf Distribution}}

\vspace{.1cm}

{\large \textbf{Marta P\'erez-Casany$^a$ and  Aina Casellas$^b$
\footnote{Address for correspondence: Marta P\'erez-Casany,
Dept. Applied mathematics II and Dama-UPC, Technical University of Catalonia
$^a$,  Barcelona, Spain. E-mail: marta.perez@upc.edu}}\\
{\large a. Department of Applied Math 2 and Dama-UPC, Technical University of
Catalonia}\\
{\large b.  Technical University of
Catalonia}

\vspace{.3cm}




\textbf{Abstract:}
The Zipf distribution also known as scale-free distribution or discrete Pareto distribution, is the particular case of  Power Law distribution with support the  strictly positive integers. It is a one-parameter distribution with a linear behaviour in the log-log scale. In this paper the  Zipfian distribution is generalized by means of the Marshall-Olkin transformation. The new model has more flexibility to adjust  the  probabilities of the first positive integer numbers  while keeping the linearity of the tail probabilities. The main properties of the new model are presented,  and several data sets  are analyzed in order to show the gain obtained by using the generalized model.

\bigskip

\vskip 0.2cm
\noindent
{{\sl{key words and phrases}}:  Scale-free distribution; Zipf distribution; Marshall-Olkin transformation; skew distributions; scale-free network.}
\bigskip


\section{Introduction}

The  Zipf distribution \cite{Zipf} appears very often in practice when modelling natural as well as man-made pehnomena.  This is because of its simplicity and its suitability to capture  the main sample characteristics. Between these characteristics one wants to enhace the following: a) large probability at one in most of its parameter space, b) long right tail, c) linearity in the log-log scale and d) scale-free. Nevertheless, in many cases  the proportion of the first few positive integer values observed,  differs considerably from the  expected probabilities under the  Zipf distribution. This is  a consequence of the fact that, in those situations, the linear behaviour  only holds for large integer values.  


\vskip 0.2cm
\noindent
The Zipf distribution is a particular case of the discrete Power Law (PL) distribution. In \cite{Clausset} it appears more that twenty  situations corresponding to different research areas where the PL distribution has been considered as a candidate distribution. The areas correspond to physics, biology, information sciences, social networking, engineering or social sciences. For example, in real world one observes that a few mega cities contain a population that is orders of magnitude larger than the mean population of  cities, and a lot of citites  have a much smaller population. In internet one observes that  very few sites contain milions of links, but many sites have just one or two links, or that millions of users visit a few select sites, giving little attention to millions of other sites (see \cite{Adamic}).

\vskip 0.2cm
\noindent
Reserchers from linguistics, ecology, demography, economy, genetics or, more recently, social networking use the Zipf distribution to model data  usually presented as rank data or frequencies of frequencies data.

\vskip 0.2cm
\noindent
The objective of the paper is to define a two-parameter generalization of the Zipf distribution that is much more flexible in modelling the probability for the first  positive integer values while at the same time allowing for both  concave and convex representations of the first probabilities in the log-log scale.  This is done by applying the transformation defined by Marshall and Olkin in 1997 \cite{MO}. In  Marshall and Olkin's original paper,  the transformation is used to generalize the exponential and the Weibull distributions. Later, the transformation has been used to generalize the Lomax, the Pareto or the Log-normal distributions. Several papers that appear in the last few years  apply  the generalizations in reliability, in time series and in  censored data. See for instance \cite{Alice}, \cite{Ghitany1, Ghitany2} or \cite{Gui}.  In \cite{Rubio} that transformation is presented as a skewing mechanism, and several classes of unimodal and symmetric distributions are extended in that manner.

\vskip 0.2cm
\noindent
The paper is organized as follows. Section 2 is devoted to the preliminars. In Section 3 the Marshall-Olkin generalized Zipf distribution is defined and its main properties are presented. In Section 4, three large  data sets from  very different research areas are analyzed  proving the usefulnes of the model presented.


\section{Preliminars}
\subsection{The Zipfian (Zipf) distribution}
A random variable (r.v.) $X$  is said to follow a Zipf distribution with {\sl{scale}} parameter $\alpha>1$ if, and only if, its probability mass function (pmf) is equal to:
\begin{equation} \label{eq:probzipfian}
P(X=x)=\frac{x^{-\alpha}}{\zeta(\alpha)},\,\, for\,\,\,\, x=1,2,3,\cdots,
\end{equation}
where $\zeta(\alpha)=\sum_{k=1}^{+\infty} k^{-\alpha}$ is the Riemann zeta function. The Zipf distribution \cite{Zipf}  is the particular case of the discrete PL distribution with support the positive integers larger than zero, and  it can also be viewed as the discretization of the Pareto (Type I) distribution. 
The Zipf distribution is  often suitable to fit  data that  correspond to frequencies of frequencies or to ranked data. These type of data show a widespread pattern in their measurements with a very large probability at one and a very small probability at some very large values. Moreover, from (\ref{eq:probzipfian}) one obtains that the  Zipf distribution will be appropiate when the data show a linear pattern in a log-log scale,   because
$$
\log (P(X=k))=-\alpha \log(x)-\log(\zeta(\alpha)).
$$
The continuous PL distribution is the only continuous distribution such that the shape of the distribution curve does not depend on the scale on which the measures are taken \cite{new}. For this reason, it and its discrete version are also konwn as {\sl{scale-free}} distributions.

\vskip 0.2cm
\noindent
Given that 
$$
F(X)=P(X \le x)=\frac{1}{\zeta (\alpha)} \sum_{k=1}^x k^{-\alpha}, 
$$
the survival or reliability function (SF) of the Zipf distribution is equal to:
\begin{equation} \label{eq:survivalzipf}
 {\overline{F}}(X)=P(X>x)=\frac{\zeta(\alpha, x+1)}{\zeta(\alpha)}.
\end{equation}
where $\zeta(\alpha, x)$ is the Hurwitz zeta function, defined as:
\begin{equation} \label{eq:xix}
\zeta(\alpha,x)=\sum_{k=x}^{+\infty} k^{-\alpha}.
\end{equation}
The mean of a Zipf distribution is finite  for $\alpha>2$, and the variance is finite only when $\alpha>3$.  Assuming that $\alpha>3$,
$$
E(X)=\frac{\zeta(\alpha-1)}{\zeta(\alpha)},\hskip 0.5cm \textrm{and} \hskip 0.5cm Var(X)=\frac{\zeta(\alpha-2) \zeta(\alpha)-(\zeta(\alpha-1))^2}{(\zeta (\alpha))^2}.
$$
\vskip 0.2cm
\noindent
Often in practice  the Zipf distribution fits well the probabilities for large values of $X$ but not the probabilities  for the smaller ones. Plotting the observed values in the log-log scale usually one observes a concave or a convex shape for the smallest values  and the linearity holds only for values larger than a given positive value. In order to avoid this problem and  increase the goodness of fit of the model, in this paper the Zipf distribution is generalized by means of the Marshall-Olkin transformation.

\subsection{The Marshall-Olkin transformation}
In 1997, Marshall-Olkin defined a method of generalizing a given  probability distribution increasing the number of parameters by one. Assume that $X$ is a r.v. with a given probability distribution with survival function ${\overline{F}} (x)$, for 
$-\infty < x < +\infty$. The Marshall-Olkin extension of the initial family is defined to be the family of distributions with survival function equal to:
\begin{equation} \label{eq:transform}
{\overline{G}} (x)=\frac{ \beta\,\, {\overline{F}} (x)}{1-{\overline{\beta}}\, {\overline{F}} (x)},\hskip 0.2cm -\infty< x < +\infty,\hskip 0.2cm \beta >0, \hskip 0.2cm \textrm{and}\,\,{\overline{\beta}}=1-\beta.
\end{equation}
The  new family contains the initial family as a particular case, obtained when $\beta=1$. The transformation proposed has an {\sl{stability property}} in the sense that the result of applying twice the transformation is also in the extended model.

\section{The Marshall-Olkin Extended Zipf Distribution}

The {\sl{Mashall-Olkin Extended Zipf}} model (MOEZipf) is defined to be the set of probability distributions with SF: 
\begin{equation} \label{eq:survivalextended}
{\overline{G}}(x; \alpha, \beta)=\frac{\beta\,\, {\overline{F}}(X)}{1-{\overline{\beta}}\,{\overline{F}}(X)}=\frac{\beta\,\,\zeta(\alpha, x+1)}{\zeta(\alpha)-{\overline{\beta}}\, \zeta(\alpha+1)},\hskip 0.3cm \text{for} \hskip 0.3cm \beta >0,\hskip 0.3cm \text{and} \hskip 0.3cm \alpha>1,
\end{equation}
where ${\overline{F}} (x)$ is the SF of the Zipf($\alpha$) distribution in (\ref{eq:survivalzipf}).
If $Y$ is a r.v. with a MOEZipf($\alpha, \beta$) distribution, then its probability mass function (pmf) is equal to:
\begin{eqnarray}\label{eq:prob}
P(Y=x) & = & {\overline{G}} (x-1; \alpha, \beta)-{\overline{G}} (x;\alpha, \beta) \nonumber \\
& = & \frac{x^{-\alpha}\,\beta\,
\zeta(\alpha)}{[\zeta(\alpha)-{\overline{\beta}} \zeta(\alpha, x)] [\zeta(\alpha)-{\overline{\beta}} 
\zeta(\alpha, x+1)]}, \hskip0.2cm  x=1,2,3,\cdots
\end{eqnarray}

In Figure \ref{fig:pmf} one can see the pmf of the MOEZipf($\alpha,\beta$) distribution for $\alpha=1.8$ and different values of $\beta$. It can be  appreciated that the probability at one increases as $\beta$ tends to zero and decreases as $\beta$ tends to infinity. This result is proved next, together  with some other results that come from comparing the probabilities of the
Zipf and the MOEZipf distributions.

\begin{figure}[tbp]
\centering
\includegraphics[angle=0,width=10cm,keepaspectratio]{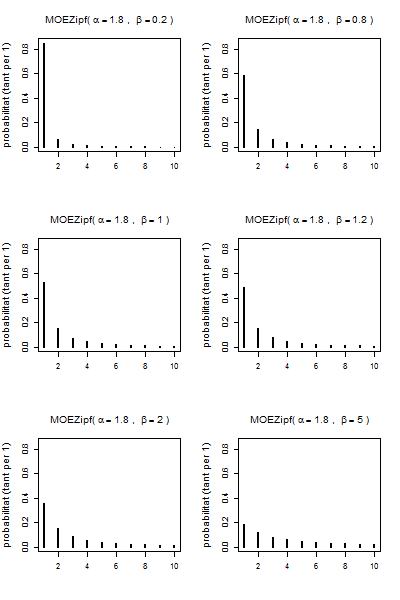}
\label{fig:pmf}
\caption{Pmf's for the MOEZipf($\alpha, \beta$) model for $\alpha=1.8$ and $\beta=0.2, 0.8, 1, 1.2, 2$ and $5$. For $\beta=1$, one obtains the Zipf(1.8) distribution.}
\end{figure}

\begin{proposition}
The probability at one of a r.v. $Y$ with a MOEZipf distribution is a decreasing function of $\beta$ verifying that:
\begin{itemize}
\item[a)] $P(Y=1)$ tends to $1$ when $\beta$ tends to  $0$
\item[b)] $P(Y=1)$ tends to $0$ when $\beta$ tends to $+ \infty$.
\end{itemize}
\end{proposition}
\vskip 0.2cm
\noindent
{\sl{Proof:}} Taking into account that by (\ref{eq:xix}),  $\zeta(\alpha,1)=\zeta(\alpha)$ and that  $\zeta(\alpha,2)=\zeta(\alpha)-1$, and setting $x=1$ in (\ref{eq:prob}), one has that:
$$
P(Y=1)=\frac{1}{1+\beta\, \zeta(\alpha,2)},
$$  
which is a decreasing function of $\beta$ that tends to zero when $\beta$ tends to infinity and to one when $\beta$ tends to zero. 
$\hfill{\square}$

\begin{proposition}
For large values of $x$, parameter $\beta$ may be interpreted as the ratio between the probabilities of a r.v. $Y$ with a MOEZipf($\alpha,\beta$) distribution and the probabilities of a  r.v. $X$ with a  Zipf($\alpha$) distribution  at $x$.
\end{proposition}
\vskip 0.2cm
\noindent
{\sl{Proof:}}
Taking into account that when $x$ tends to infinity $\zeta(\alpha, x)$  tends to zero, one has that:
$$
\lim_{x\to +\infty} \frac{P(Y=x)}{P(X=x)}=\lim_{x\to +\infty}\frac{\beta (\zeta(\alpha))^2}{ [\zeta(\alpha)-{\overline{\beta}} \zeta (\alpha, x)] [\zeta(\alpha)-{\overline{\beta}} \zeta (\alpha, x+1)] }=\beta.
$$
$\hfill{\square}$

\begin{proposition}
Let $Y$ be a r.v. with a MOEZipf($\alpha$, $\beta$) distribution and $X$ be a r.v. with a Zipf($\alpha$) distribution.
For any $x\ge 1$ one has that: if $\beta=1$, $P(Y=x)=P(X=x)$, if $\beta>1$, $\beta \, P(Y=x)\ge P(X=x)$, and if $\beta<1$, $P(Y=x)\ge \beta\, P(X=x)$.
\end{proposition}
\vskip 0.2cm
\noindent
{\sl{Proof:}}
\begin{itemize}
\item If $\beta=1$, the probabilities in (\ref{eq:prob}) correspond to those of the Zipf($\alpha$) distribution.
\item If $\beta>1$, $\overline{\beta}<0$, given that for all $x\ge 1$ $\zeta(\alpha, x)\le \zeta (\alpha)$, one has that $\overline{\beta} \zeta(\alpha, x)\ge \overline{\beta} \zeta (\alpha)$. Thus,
\begin{eqnarray*}
0 \le \zeta(\alpha)-\overline{\beta} \zeta(\alpha, x)\le \beta \zeta(\alpha) &  \Leftrightarrow & 
0 \le [\zeta(\alpha)-\overline{\beta} \zeta(\alpha, x)] [\zeta(\alpha)-\overline{\beta} \zeta(\alpha, x+1)]\le (\beta\, \zeta(\alpha))^2 \\ & \Leftrightarrow &
\frac{x^{-\alpha}\,\beta\,
\zeta(\alpha)}{[\zeta(\alpha)-{\overline{\beta}} \zeta(\alpha, x)] [\zeta(\alpha)-{\overline{\beta}} 
\zeta(\alpha, x+1)]} \ge \frac{x^{-\alpha}}{\zeta(\alpha)} \frac{1}{\beta}\\ & \Leftrightarrow & P(Y=x)\ge P(X=x) \frac{1}{\beta}.
\end{eqnarray*}
\item If $\beta <1$, $\overline{\beta}>0$, then $\overline{\beta} \zeta(\alpha, x)\le \overline{\beta} \zeta (\alpha)$. Thus, for any 
$x \ge 1$,
$0 \le \zeta(\alpha)-{\overline{\beta}} \zeta(\alpha, x) \le \zeta (\alpha)$, which gives that:
\begin{eqnarray*}
(\zeta(\alpha))^2 & \ge & [\zeta(\alpha)-{\overline{\beta}} \zeta(\alpha, x)] [\zeta(\alpha)-{\overline{\beta}} \zeta(\alpha, x+1)] \\ & \Leftrightarrow & \frac{x^{-\alpha} \beta}{\zeta(\alpha)} \le \frac{x^{-\alpha} \beta \zeta(\alpha)}{[\zeta(\alpha)-{\overline{\beta}} \zeta(\alpha, x)] [\zeta(\alpha)-{\overline{\beta}} \zeta(\alpha, x+1)]} \\ & \Leftrightarrow &  \beta P(X=x) \le P(Y=x).
\end{eqnarray*}
\end{itemize}
$\hfill{\square}$

\vskip 0.2cm
The MOEZipf distribution is only linear in the log-log scale for large values of $x$ as is proved in the following result:

\begin{proposition}
Let $Y$ be a r.v. with a MOEZipf($\alpha, \beta$) distribution. For large values of $x$, $\log(P(Y=x))$ is a linear function of $\log(x)$. 
\end{proposition}
\vskip 0.2cm
\noindent
{\sl{Proof:}} The proof is a straightforward consequence of the fact that (\ref{eq:xix}) tends to zero when $x$ tends to infinity.   Hence, for large values of $x$, the denominator of (\ref{eq:prob}) is approximatively  equal to $\zeta(\alpha)^2$, and thus one has that:
\begin{equation} \label{eq:linear}
\log(P(Y=x))\simeq -\alpha \log (x)+\log (\beta)-\log(\zeta(\alpha)).
\end{equation}
$\hfill{\square}$

\vskip 0.2cm
\noindent{{REMARK:}} The  result is also true  in a larger support of the distribution if $\alpha$ is large. This is because (\ref{eq:xix}) is very small for large values of $\alpha$ if $x\ge 2$.

\vskip 0.2cm
\noindent
Figure \ref{fig:linearity} shows the behaviour of the MOEZipf distribution in the log-log scale, for different parameter values, together with the straight line obtained by changing $\simeq$ by $=$ in (\ref{eq:linear}).
\begin{figure}[htbp] \label{fig:linearity}
\begin{center}
\includegraphics[angle=90,width=15cm,keepaspectratio]{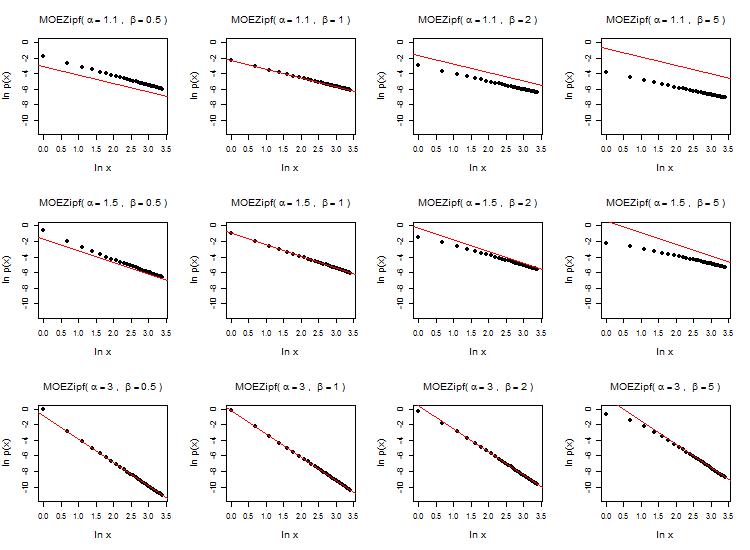}
\caption{Probabilities of the MOEZipf($\alpha, \beta$) distribution in the log-log scale. Each row corresponds to a different value of $\alpha$ and each column to a different value of $\beta$. The parameter values considered are: $\alpha=1.1, 1.5$ and $3$ and $\beta=0.5, 1, 2$ and $5$.}
\end{center}
\end{figure}

\vskip 0.2cm
Next proposition compares the ratio of two consecutive probabilities of a MOEZipf and a Zipf distributions with the same  
$\alpha$ value. As it can be appreciated in Figure \ref{fig:ratioprob}, the ratio only shows important differences for small values of $x$. Moreover, for values of $\beta$ smaller (larger) than one, the ratios corresponding to the MOEZipf distribution are smaller (larger) than the ratios associated to the Zipf distribution. This result is stated in the next proposition. 

\begin{proposition} If $Y$ is a r.v. with a MOEZipf($\alpha$, $\beta$) distribution and $X$ is a r.v. with a Zipf($\alpha$) distribution, one has that:
\begin{itemize}
\item[1)] If $\beta > (<)\, 1$, then
$$
\frac{P(Y=x+1)}{P(Y=x)} > (<)\, \frac{P(X=x+1)}{P(X=x)}.
$$
\item[2)] $\forall\, \beta >0$,
$$
 \lim_{x\to +\infty} \frac{P(Y=x+1)}{P(Y=x)}=\frac{P(X=x+1)}{P(X=x)}.
$$
\end{itemize}
\end{proposition}
\vskip 0.2cm
\noindent
{\sl{Proof:}} By (\ref{eq:prob}) one has that
\begin{eqnarray} \label{eq:pk+1/pk}
\frac{P(Y=x+1)}{P(Y=x)} & = & \Big(\frac{x}{x+1}\Big)^{\alpha} \,\, \frac{\zeta(\alpha)-\overline{\beta}\, \zeta(\alpha, x)}{\zeta(\alpha)-\overline{\beta}\,\zeta(\alpha, x+2)} \nonumber \\ 
 & = & \frac{P(X=x+1)}{P(X=x)}\,\, \frac{\zeta(\alpha)-\overline{\beta}\, \zeta(\alpha, x)}{\zeta(\alpha)-\overline{\beta}\,\zeta(\alpha, x+2)}.
\end{eqnarray}
Given that for any $\alpha>1$ and any $x\ge 1$, $\zeta(\alpha, x+2) < \zeta (\alpha, x)$, it is possible to state that if  $0< \beta < 1$
($\beta >1$),
$$
1 > (<) \frac{\zeta(\alpha)-(1-\beta) \zeta (\alpha, x)}{\zeta(\alpha)-(1-\beta) \zeta (\alpha,x+2)},
$$
which proves point $1)$. Given that (\ref{eq:xix}) tends to zero when $x$ tends to infinity, point $2)$ is a straight consequence of
(\ref{eq:pk+1/pk}). {\hfill{$\square$}}

\begin{figure}[htbp]
\begin{center} \label{fig:ratioprob}
\includegraphics[angle=0,width=13cm,keepaspectratio]{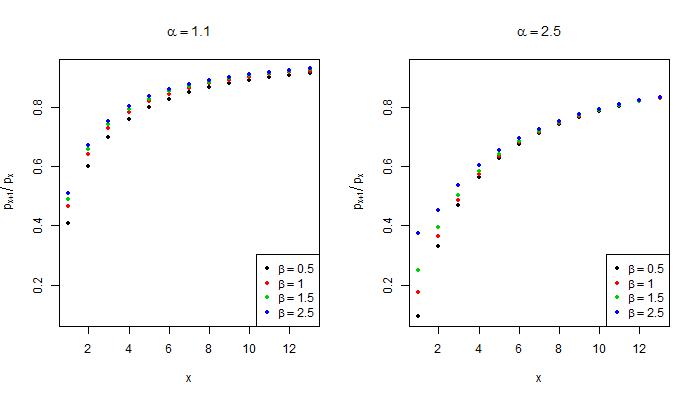}
\caption{Ratio of two consecutive probabilities for the MOEZipf($\alpha,\beta$) and Zipf($\alpha$) distributions with $\alpha$ equal to $1.1$ and $2.5$ and $\beta$ equal to $0.5, 1, 1.5$ and $2.5$.}
\end{center}
\end{figure}

\subsection{Parameter estimation}

In this subsection two ways of estimating the parameters of the  MOEZipf($\alpha, \beta$) distribution are considered. The  maximum likelihood estimator (m.l.e) denoted by $(\hat \alpha, \hat \beta)$, is obtained by maximizing
the corresponding log-likelihood function, that in that case takes the form:
\begin{eqnarray*}
l(\alpha, \beta; y_i) & = & n \log(\beta)+n \log(\zeta(\alpha))-\alpha \sum_{i=1}^n \log(y_i)-\sum_{i=1}^n log(\zeta(\alpha)-\overline{\beta} \zeta(\alpha, x))\\
& - & \sum_{i=1}^n \log(\zeta(\alpha)-\overline{\beta} \zeta (\alpha, x+1)).
\end{eqnarray*}

\vskip 0.2cm
\noindent
The second method of estimation considered consists of solving numerically the system of equations that comes from equating the observed and expected probabilities at one, and the sample mean to the expected value of the distribution. If one denotes by $f_1$ the observed frequency at one, and by $\overline{y}$ the sample mean, that leads to solving
\begin{equation} \label{eq:sistema}
\begin{cases} (\zeta(\alpha, 2)\, \beta+1)^{-1}=\frac{f_1}{n} & \text{}
\\
E(Y)=\overline{y},
\end{cases}
\end{equation}
which can be done by first  solving the following equation in $\alpha$:
$$
\frac{n-f_1}{f_1\,\cdot \zeta(\alpha,2)} \zeta(\alpha)\sum_{x=1}^{\infty} \frac{x^{-\alpha+1}}{[\zeta(\alpha)-\frac{f_1\,\cdot \zeta(\alpha)-n}{f_1\, \cdot\zeta(\alpha,2)} \zeta(\alpha, x)][\zeta(\alpha)-\frac{f_1\cdot\zeta(\alpha)-n}{f_1\,\cdot \zeta(\alpha,2)} \zeta(\alpha, x+1)]}=\overline{y},
$$
and then estimating $\beta$ by substituing the estimation of $\alpha$   in the first equation of (\ref{eq:sistema}).
The solution obtained by this method is denoted by $(\tilde \alpha, \tilde \beta)$.

\section{Data analysis}

In this section three sets of data are fitted by means of the Zipf and the MOEZipf models, and the results obtained are compared. All the data sets have a very large sample size and correspond to real data obtained from different areas.

\subsection{Example from Linguistics}

The data of this example corresponds to the  frequency of occurrence of words in the novel {\sl{Moby Dick}} by Herman Melville and can be found in: 

http://tuvalu.santafe.edu/~aaronc/powerlaws/data.htm. 

\vskip 0.2cm
This set of data was first analyzed in \cite{Zipf} which  is the reference where the Zipf distribution was defined. More recently \cite{Ferrer}  and \cite{Clausset}  have also considered  this set of data. The first ones use the data set to compare real with random texts, and the second ones fit the data by means of a general PL distribution. The set contains the frequencies of a total of $18855$ words and nearly $75\%$ of the observations correspond to the first three positive integer values. The three most frequent words appear $14086$, $6414$ and $6260$ times. The observations larger than or equal to $53$ have been grouped, in order to be able to compare the two models by means of the $\chi^2$ goodness-of-fit statistic.

\vskip 0.2cm
\noindent
The m.l.e estimations of $\alpha$ obtained assuming the two different models do not differ considerably. Nevertheless, the m.l.e of parameter $\beta$ under the MOEZipf model is $50\%$ larger than under the Zipf model ($\beta=1$). The reduction obtained in the $\chi^2$ Pearson statistic using the proposed model instead of the Zipf model is equal to $79.64\%$. The AIC as well as the log-likelihood  show that the generalized model estimating the parameters by maximum likelihood gives the best fit. In Figure \ref{fig:words} one can see the data together with the two fitted models in a log-log scale.

\begin{table}[h]
\begin{center}
\begin{tabular}{lcccccc}
  \noalign{\hrule height 1pt}
   Distrib. & Param. &  Estimat. &   log-like. & $X^2$ & p-val. &   AIC \\ 
   \noalign{\hrule height 1pt}
      Zipf  &  $\hat{\alpha}$  &   1.775  &  -40196.00 &  272.38 &  0 & 80394.00\\
   \hline
   MOEZipf  &  $\tilde{\alpha}$  & 1.908  &  -40086.28 & 62.96 &  0.097 & 80176.56\\
   (2nd method)         &  $\tilde{\beta}$   &   1.429 & & & & \\
   \hline
   MOEZipf  &  $\hat{\alpha}$  &   1.944  &  \textbf{-40082.42} &  \textbf{55.45}  & 0.293  & \textbf{80168.83}\\
    (m.l.e)        &  $\hat{\beta}$   &   1.523 & & & & \\
  \noalign{\hrule height 1pt}
\end{tabular}
\caption{Reults of fitting the variable: {\emph{Frequency of occurence of words}}, in \emph{Moby Dick}.}
\label{sum_ajust_words}
\end{center}
\end{table}

\begin{figure}[tbp]
\begin{center} \label{fig:words}
\includegraphics[angle=0,width=9cm,keepaspectratio]{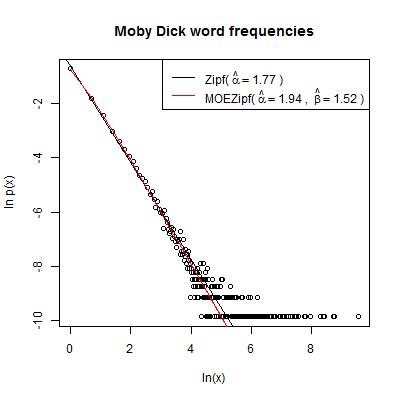}
\caption{Observed and expected data in the log-log scale, for the word frequency data.}
\end{center}
\end{figure}

\subsection{Example from electronic mail}

Given a database containing different electronic mail addresses, one can count how many connections one address has had in a given  period of time. The table of frequencies of such a r.v. tends to have large probability at one  (most of the addresses only have one contact), and a very small probability at some large values (just few addresses have lots of contacts). The data set analyzed in this example corresponds to the number of connections of a total of $225409$ electronic addresses, and may be found in http://snap.stanford.edu/data/email-EuAll.html. They were collected between october 2003 and may 2005. In \cite{Leskovec} this  data set is  analyzed by fitting a PL distribution in  the tail of the distribution.

\vskip 0.2cm
\noindent
Here $85\%$ of the observations are equal to one. The observed probabilities of the first  five values decrease very quickly and after these values, they decrease more slowly. The three addresses with the largest number of contacts have exactly $930$, $871$ and  $854$ contacts.  After grouping the data larger than $65$,   the $\chi^2$ statistic is reduced in a $93.74\%$ by using the generalized model instead of the original by m.l. The  AIC criterium concludes that the MOEZipf model is the best one.

\begin{table}[ht]
\begin{center}
\begin{tabular}{lcccccc}
  \noalign{\hrule height 1pt}
   Distrib. & Param. &  Estimat.  &   log-like. & $X^2$ & p-val. &   AIC \\ 
   \noalign{\hrule height 1pt}
      Zipf  &  $\hat{\alpha}$  &   2.968  &  -156765.21 &  13714.84 &  0 & 313532.42\\ 
   \hline
   MOEZipf  &  $\tilde{\alpha}$  & 2.126  &  -154526.75 & 968.34 &  0 & 309057.51\\
    (2nd method) &  $\tilde{\beta}$   &   0.321 & & & & \\ 
   \hline
   MOEZipf  &  $\hat{\alpha}$  &   2.284  &  \textbf{-154399.82} &  \textbf{858.27}  &  0 & \textbf{308803.64}\\
     (m.l.e.)       &  $\hat{\beta}$   &   0.390 & & & & \\
  \noalign{\hrule height 1pt}
\end{tabular}
\caption{Results of fitting the variable: \emph{Number of relations by electronic mail}. }
\label{sum_ajust_email}
\end{center}
\end{table}
  
\vskip 0.2cm
\noindent
Figure \ref{fig:mail} shows the observed and fitted probabilities in the log-log scale. It can be appreciated that the convex behaviour of the MOEZipf model gives place to a better fit, not only for the first values of the distribution but also for the values in the tail.

\begin{figure}[tbp]
\begin{center} \label{fig:mail}
\includegraphics[angle=0,width=9cm,keepaspectratio]{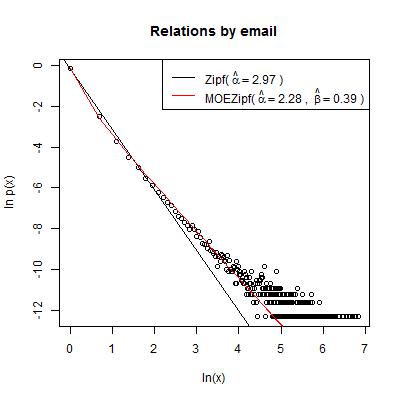}
\caption{Observed and expected data in the log-log scale, for the e-mail example.}
\end{center}
\end{figure}

\subsection{Example from citations}

The last example considered corresponds to the number of times that a given paper is cited in a given database. This is an  important variable because it allows one to calculate the impact factor of the scientific journals. The database analyzed has a total of $32158$ papers in the area of {\sl{High-energy physics}}, published in arXiv.org between January  1993 and April 2003,
and may be found in http://snap.stanford.edu/data/citHepPh.html. This data set has also been analyzed in  \cite{Leskovec}. 

\vskip 0.2cm
\noindent
From that data one observes that the $26\%$ of the probability corresponds to the  first two values, meaning that more than a quarter of the papers are cited  at most twice.  For model fitting, we have grouped the values larger than $119$.  As in the  previous examples, Table \ref{sum_ajust_citesReb} indicates that 
the MOEZipf model estimating  by m. l. provides the best fit. The inclusion of the $\beta$ parameter implies a reduction of a $87.73\%$ in the $\chi^2$ statistic. It is important to note that  $\hat \beta$ is equal to $13.1$ which is a very large value, compared with the value of $1$ that corresponds to the Zipf distribution. In Figure \ref{fig:cites} it is possible to appreciate that the MOEZipf model shows a concave behaviour that improves considerably the fit of the first values as well as the one in the tail of the distribution.  

\begin{table}[!htb]
\begin{center}
\begin{tabular}{lcccccc}
  \noalign{\hrule height 1pt}
   Distrib. & Param. &  Estim.  &   log-like & $X^2$ & p-val. &   AIC \\ 
   \noalign{\hrule height 1pt}
      Zipf  &  $\hat{\alpha}$  &   1.421  &  -105839.81 &  13172.05 &  0 & 211681.61\\ 
   \hline
   MOEZipf  &  $\tilde{\alpha}$  & 2.214  &  -99490.01 & 1615.25 &  0 & 198984.03\\
   (2nd met.) &  $\tilde{\beta}$   &   11.677 & & & & \\ 
   \hline
   MOEZipf  &  $\hat{\alpha}$  &  2.161   &  \textbf{-99197.93} &  \textbf{816.62}  &  0 & \textbf{198399.87}\\
    (m.l.e) &  $\hat{\beta}$   &  13.058  & & & & \\
  \noalign{\hrule height 1pt}
\end{tabular}
\caption{Results of fitting the r.v.: \emph{Number of citations of a given paper}.}
\label{sum_ajust_citesReb}
\end{center}
\end{table}

\begin{figure}[htbp]
\begin{center} \label{fig:cites}
\includegraphics[angle=0,width=9cm,keepaspectratio]{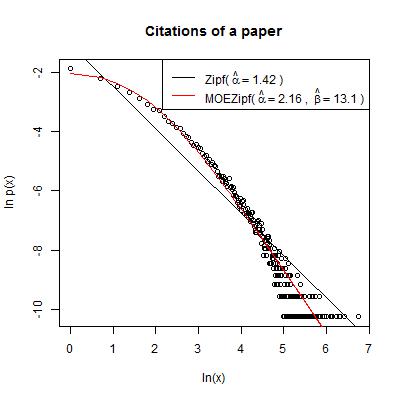}
\caption{Observed and expected data in the log-log scale, for the citations example.}
\end{center}
\end{figure}

\section{Conclusions}

The Marshall-Olkin transformation has proved to be useful for generalizing the Zipf distribution both in terms of providing good properties as well as in terms of improving the goodness of fit obtained in the data sets analyzed. The extended model can show concavity or convexity in the first part of the domain, as it is shown by means of the examples.
The linear behaviour is always observed in the tail of the distribution. 
The extra-parameter also allows for ratios between two consecutive probabilities  larger or smaller than the corresponding ratio of a Zipf distribution. In the three data sets  considered, the fittings obtained for the first values are considerably better than the ones corresponding to the Zipf distribution, but  they are also better in the tail.  The reduction in the $\chi^2$ goodness-of-fit statistic has always been larger or equal than $80\%$. The AIC   points out the MOEZipf model as the better model in all the 
examples considered.  

\vskip 0.3cm
\noindent
{\bf{Acknowledgements.}} The authors want to thank D. Dominguez-Sal and J.LL. Larriba-Pey for their help in providing the data sets that are analyzed in this work and to J. Ginebra for his interesting comments and suggestions that helped to considerably improve the manuscript. The first author also wants to thank the Spanish Ministery of Science and Innovation for  Grants. No. TIN2009-14560 and MTM-2010-14887, and Generalitat de Catalunya for Grant No. SGR-1187.

\end{document}